\documentclass[12pt]{article}

\usepackage[dvips]{graphicx}

\begin{document}


\noindent{\bf DISLOCATION MOBILITY IN TWO-DIMENSIONAL LENNARD-JONES MATERIAL}

\vskip 10 pt
\noindent{N. P. BAILEY*, J. P. SETHNA* AND C. R. MYERS**}

\vskip 10 pt
\noindent{* Physics Department, Cornell University, 117 Clark Hall, Ithaca, NY 14853}

\noindent {** Cornell Theory Center, Cornell University, Ithaca, NY 14853}


\vskip 15 pt
\noindent{\bf ABSTRACT}
\vskip 10 pt

In seeking to understand at a microscopic level the response of dislocations to stress we have undertaken to study as completely as possible the simplest case: a single dislocation in a two dimensional crystal. The intention is that results from this study will be used as input parameters in larger length scale simulations involving many defects. We present atomistic simulations of defect motion in a two-dimensional material consisting of atoms interacting through a modified Lennard-Jones potential. We focus on the regime where the shear stress is smaller than its critical value, where there is a finite energy barrier for the dislocation to hop one lattice spacing. In this regime motion of the dislocation will occur as single hops through thermal activation over the barrier. Accurate knowledge of the barrier height is crucial for obtaining the rates of such processes. We have calculated the energy barrier as a function of two components of the stress tensor in a small system, and have obtained good fits to a functional form with only a few adjustable parameters.

\vskip 15 pt
\noindent{\bf INTRODUCTION}
\vskip 10 pt

This paper is concerned with the motion of a single dislocation. Thus there is no dislocation-dislocation interaction; the interaction is between the dislocation and the applied stress. Furthermore we work in two dimensions, which eases the computational burden and aids visualization. Having simplified the problem to this extent, we have a chance of understanding it in detail. Once such understanding is developed, it will then make sense to proceed to more realistic, though computationally more expensive, cases (e.g. three dimensions, realistic potentials, etc.).

Our system consists of a relatively small ($<100$) number of atoms in two dimensions (2D) interacting through a Lennard-Jones potential which has been truncated and made to go smoothly to zero at the cutoff distance ($2.7\sigma$; this is large enough for third neighbor interactions to be included). The system has periodic boundary conditions in the vertical direction, and rigid \lq walls' on the sides. The walls are simply lines of atoms which are constrained to move as rigid bodies. In addition the atoms in each of the next-to-outermost columns are constrained to more rigidly in the $x$-direction, and independently in the $y$-direction. This system for the boundaries is due to Tomasi \cite{tomasi}. If a shear stress is applied to the boundary walls the dislocation will move by glide, but only if the shear stress is above a certain critical value $\sigma_c$. The applied shear stress is the resolved shear stress in this geometry. Note that the critical resolved shear stress for dislocation motion depends on the other components of stress, hence knowledge of the resolved shear stress alone is not enough to decide whether a given dislocation will move or not. At zero temperature, with $\sigma_{xy} < \sigma_c$, the dislocation cannot move, but with a finite temperature, motion still occurs as thermally activated hops over an energy barrier. This barrier corresponds to the Peierls barrier for an edge dislocation in three dimensions. Our task was to calculate this barrier as a function of all three components of the stress tensor ($\sigma_{xx}, \sigma_{xy}, \sigma_{yy}$). However so far, we have only dealt with the dependence on the first two, since we have only begun to incorporate the techniques necessary for applying a constant stress in the direction in which periodic boundary conditions are imposed. We have considered only one size of system; finite size effects are important. Extensions to all three components of stress and extrapolations to large sizes are in progress.

\vskip 15 pt
\noindent{\bf THEORY}
\vskip 10 pt

\noindent\underline{Thermal Activation}

We concentrate on calculating the energy barrier to hopping for shear stress less than the critical shear stress. For these values of shear, there exist so called fixed points of the dynamics. These are associated with local minima in the potential. Two nearby minima are separated by a barrier in the energy landscape (note that the saddle point of the barrier is also a fixed point, albeit an unstable one). When the energy barrier is large compared to the temperature, the transition rate between the states will have the form

\begin{equation}
R = \nu \exp (-{{E_B}\over{k_B T}})
\end{equation}

\noindent where $E_B$ is the barrier height and $\nu$ is an attempt frequency which can be calculated from the curvature of the potential landscape near the minimum and near the barrier top \cite{rates}. Because of the exponential, however, the rate is much more sensitive to $E_B$ than it is to $\nu$. Hence it is importance to know $E_B$ accurately to be able to reasonably calculate such rates. For the dislocation, the rate of hopping is proportional to the velocity, and hence its mobility. The barrier height is defined as follows. For any path between the two minima we find the point along the path where the potential energy is greatest; call this value $E_{max}$. We then consider all paths and take the one whose $E_{max}$ is smallest. This is the {\it minimum energy path}. The barrier height is $min_{paths} \{ E_{max}\} - E_{intial state}$. The location of the maximum energy along the minimum energy path corresponds to a saddle point in the potential landscape. Several methods exist for finding barrier heights. We use one which finds the whole minimum energy path, called the \lq Nudged Elastic Band' method \cite{NEB}.

\vskip 10pt
\noindent\underline{Model and Potential}

We model a small piece of two-dimensional material containing a single edge dislocation atomistically, using methods of molecular dynamics. We use a classical pair potential defined as follows: Lennard-Jones (6-12, with standard parameters $\epsilon$ and $\sigma$) for $r < r_{cut1} = 2.41308788\sigma$, a quadratic in $r^2$ for $r_{cut1} < r < r_{cut2} = 2.7\sigma$, and zero for $r> r_{cut2}$. This potential was formulated by Chen \cite{xichen}. It is continuous and smooth everywhere. Extension to other potentials and other forms of the cutoff is planned. The units for the simulation are determined by the parameters $\epsilon$ and $\sigma$ in the potential, which are set to unity, hence all energies are in units of $\epsilon$, and distances in units of $\sigma$. Units of time, stress etc. all follow from these. Often one makes a connection with physical systems by matching the parameters to those of Argon, for which Lennard-Jones is a good potential. So, $\epsilon = 119.8 K k_B$ or $\sim 0.01$eV, and $\sigma = 0.341$nm \cite{Argon}. However since we have a 2D system, there is little useful quantitative comparison to made with experiment (for one thing, stress has different units in 2D than in 3D).

\vskip 15 pt
\noindent{\bf SIMULATION}
\vskip 10 pt

\begin{figure}[h]  
\begin{center}
\includegraphics[width=2.0in]{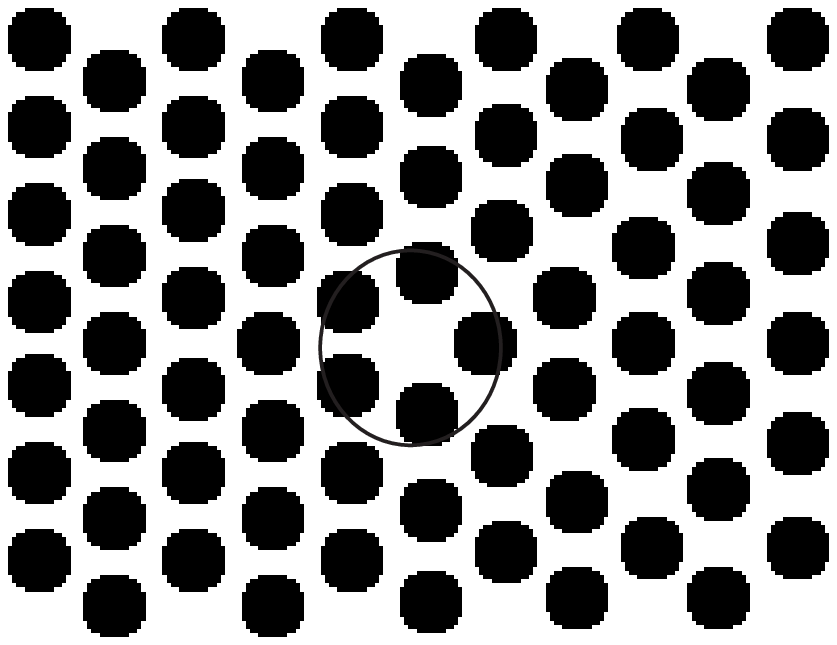}
\includegraphics[width=2.0in]{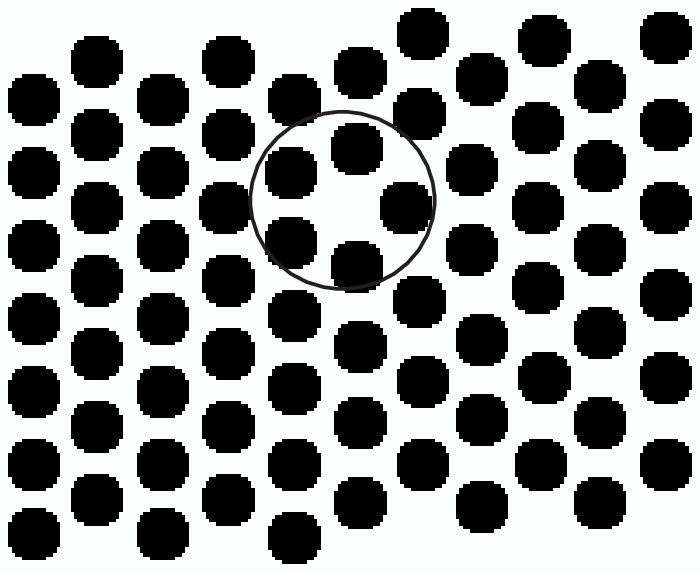}
\includegraphics[width=2.0in]{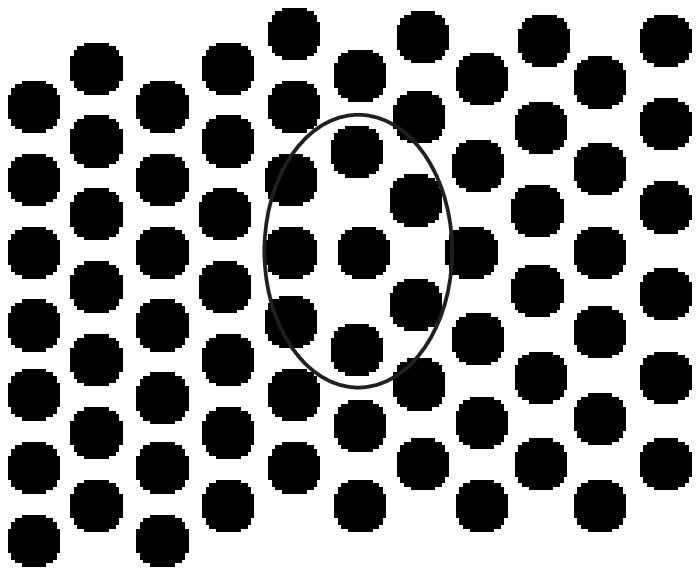}
\caption{Initial, final and saddle point configurations. These are for zero shear stress, boundaries fixed in $x$-direction. Circles indicate the dislocation.}
\label{StartFinish}
\end{center}
\end{figure}

We simulate an \lq$N \times N$' system, where $N$ is the number of rows on the left of the dislocation; there are $N-1$ on the right. Two extra columns are added to each side to form the boundaries. Typically $N=7$, which corresponds to 71 atoms. Simulation runs consist of the following procedure: A set of atoms is configured as two lattices of slightly differing lattice constants, and correspondingly different numbers of rows, placed together. The atoms are relaxed by evolving the system using Langevin dynamics, after which there was a localized dislocation. Next, a shear stress is applied which is just of sufficient strength and duration to cause the dislocation to move one lattice spacing, after which the stress is reset to zero and further relaxation is done. Copies are made of the relaxed system before and after the move. The main part of the simulation consists of a loop in which stress is applied to these copies, their energy is minimized using the \lq MDmin' procedure, taken from Ref. \cite{schiotz}, and they are passed to the  barrier finding routine, which uses the Nudged Elastic Band method \cite{NEB}. In this method a chain of replicas of the system is created forming a line in configuration space between the local minima. Forces from the potential, and between replicas are applied, with certain corrections, and the whole chain relaxed until it lies along the minimum energy path. Once the barrier is calculated, the stress is incremented and the loop repeats, minimizing the two copies now with a different stress, and so on. Fig. \ref{StartFinish} shows initial, final and saddlepoint configurations of the $7 \times 7$ system.

\vskip 10 pt
\noindent\underline{Curve-fitting: Finding the top.}

\begin{figure}
\begin{center}
\includegraphics[width=3.5in]{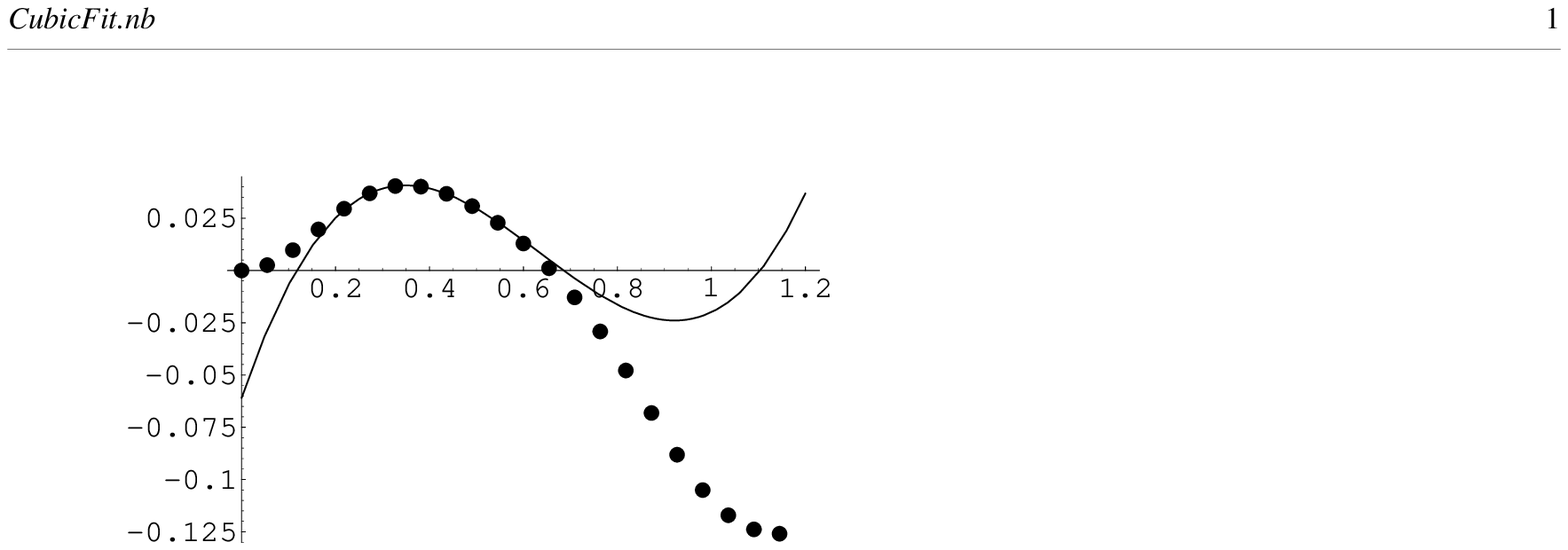}
\caption{A single barrier profile with the fitted cubic. $\sigma_{xx} = 9.6$, $\sigma_{xy} = 0.8$
}
\label{cubicFit}
\end{center}
\end{figure}

In general there will not be a replica right at the saddle point, so given the energies of the replicas it is necessary to do curve fitting to find the actual maximum energy along the path, and hence the barrier height itself. The information returned from the routine includes the configurations of the replicas and the Euclidean distance along the chain for each one, called the reaction coordinate, as well as the energies. We must subtract the work done by the external stress to find the relevant energy-quantity: at finite temperature it would be the Gibbs free energy; at zero temperature it is equal to the enthalpy, and is the quantity that is minimized in equilibrium when a constant force is applied. The set of distance-enthalpy points can be plotted in order to visualize the shape of the barrier. To find the height, a cubic is fitted to the top four points, see Fig \ref{cubicFit}; the position of the maximum can then be simply calculated. An estimate of the uncertainty can be got from considering the top five points and fitting to a quartic, and taking the difference of the two results. The difference appeared only in the fifth digit, though close to critical shear, where convergence was not as good, the relative error became large.

\vskip 10 pt
\noindent\underline{Constant pressure simulations}
	
The first runs were done with the boundary walls free only to move in the $y$-direction, and fixed $L_y$, though it was intended that we would eventually have constant $\sigma_{xx}$ and constant $\sigma_{yy}$. A material is generally under fixed stress not fixed strain, so incorporating conditions of constant stress, and hence fluctuating boundaries, is a more realistic approach. In fact we found direct evidence of the necessity for constant stress simulations when we looked at the system-size dependence of our results. The height of the barrier for $\sigma_{xy} = 0$ depended strongly on system size---it decreased by a factor of two upon going from a $7 \times 7$ system to a $13 \times 13$ one. Since we would like to believe that we can in fact get away with simulating such small systems this was a worrying fact. 

Including constant $\sigma_{xx}$ was straightforward: we let the boundary walls move and put a force in the $x$-direction on them. When the system was relaxed in the initial stage of the simulation, the final separation of the boundaries was about half a lattice constant larger than the fixed separation we had been using---the dislocation liked to take up more space than an uninterrupted column of atoms, and hence there was a significant sideways pressure in the fixed boundary simulations. When the system size was increased this pressure decreased and since it was already clear that the barrier height depended on sideways pressure, this would explain the dependence on system size in the earlier simulations. So increasing the system size at fixed $\sigma_{xx}$ should have a much smaller effect on the barrier. 

In fact the barrier height was an {\it increasing} function of the system size when $\sigma_{xx}$ was held fixed. This was thought to be due to the vertical dimension of the sample being held fixed (the dimension in which periodic boundary conditions were operating), hence there was a varying effective pressure in this direction. Since there is no rigid boundary here as on the sides, a more sophisticated technique must be used, similar to Parrinello-Rahman dynamics \cite{parrinello}. Here, the lengths of the simulation cell in the different directions are allowed to vary dynamically. The equations of motion are suitably modified to include this extra degree of freedom. In the present case this only had to be done for the vertical direction, with the variable $L_y$ being introduced. However there are subtleties associated with combining this technique with Nudged Elastic Band, not least the issue of defining an angle in a space which has one axis corresponding to a length and the rest to dimensionless positions.

\vskip 15 pt
\noindent{\bf RESULTS}
\vskip 10 pt

\begin{figure}
\begin{center}
\includegraphics[width=4in]{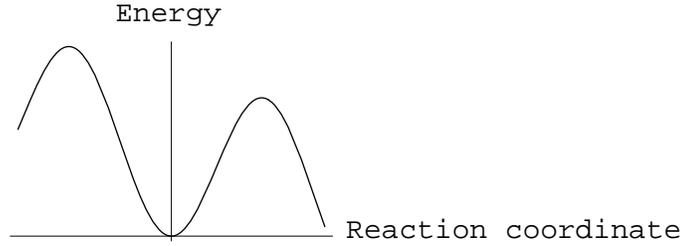}
\caption{Schematic representation of barriers in a periodic system. The shear stress is positive, hence the system wants to move to the right. However we can calculate the barrier to hop to the left by repeating the calculation with the opposite sign of shear.}
\label{barrierPicture}
\end{center}
\end{figure}

\begin{figure}[h]  
\begin{center}
\includegraphics[width=4.5in]{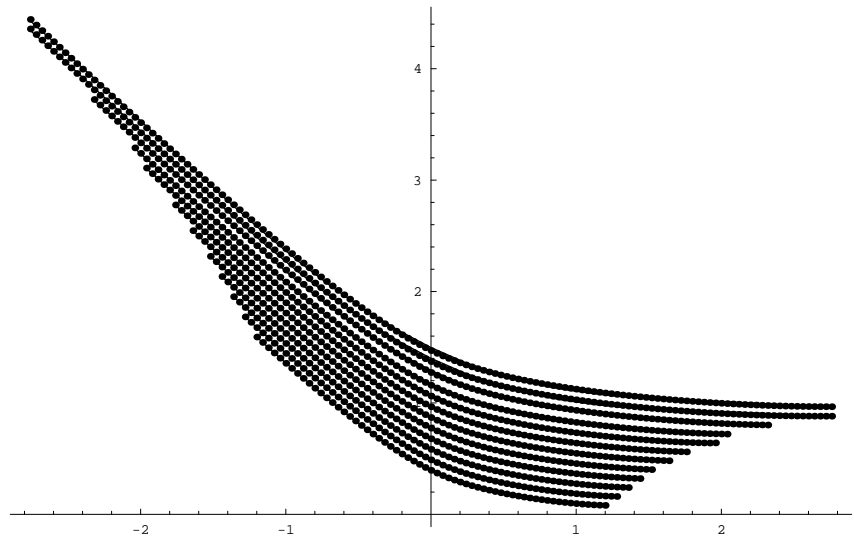}
\caption{Energy barrier versus shear stress for $8.8 < \sigma_{xx} < 13.6$. The curves have been displaced vertically for clarity. Also included, but not readily visible are the one-dimensional fits to the data.}
\label{ManyGraphs}
\end{center}
\end{figure}

The energy barrier as a function of shear stress $\sigma_{xy}$ with fixed $\sigma_{xx}$, for several values of $\sigma_{xx}$  is shown in Fig \ref{ManyGraphs}. The dots are data points from barrier calculations; the solid lines are three-parameter fits to a series expansion obtained by considering the one-dimensional barrier problem. For stress larger than the critical value, there is no fixed point, and the defect slides with periodically varying velocity. For stress smaller than the critical value, there are two fixed points, a stable one corresponding to the local minimum, and an unstable one corresponding to the barrier top (of course there are many more really, due to the periodicity of the lattice). The appearance of two fixed points as the stress goes below $\sigma_c$ (or equivalently their disappearance as stress goes above $\sigma_c$) is a {\it saddle-node bifurcation}. Note that we also have points for negative shear; this corresponds to hopping in the opposite direction for positive shear, see Fig. \ref{barrierPicture}. For large negative shear the barrier becomes simply the energy difference between the two minima. These local minima only exist for $ |\sigma_{xy}| < \sigma_c$ (beyond which the dislocation starts to slide in the appropriate direction), hence our data covers the range $-\sigma_c$ to $\sigma_c$, for several values of $\sigma_{xx}$.

\vskip 15 pt
\noindent{\bf CONCLUSIONS}
\vskip 10 pt

We have calculated the barrier height for dislocation hopping for range of both $\sigma_{xx}$ and $\sigma_{xy}$. We have shown how this data can be parametrized to reasonable accuracy with a fit that needs only a few parameters. These parameters could be used for example in a simulation whose primitive objects are dislocations which move in a stress field. The present results, augmented by $\sigma_{xx}$ dependence, could be used to calculate the motion of each dislocation given the values of the components of the stress tensor at its location (the stress field would be calculated from the elastic fields of the other dislocations plus any external sources of stress).

The next development in the simulation will be to include the dependence on the third component of the stress tensor, $\sigma_{yy}$, which corresponds to pressure on the top and bottom of the system. We have just started to be able to do runs at constant $\sigma_{yy}$, though there are subtleties in combining this with the Nudged Elastic Band method. Once we are satisfied that we can do this well, we will repeat with different inter-atomic potentials. Another aspect of simulating this system is to consider shear stress above $\sigma_c$, where the dislocation slides continuously, if not quite steadily. Preliminary studies indicate interesting behaviour, including a delocalization of the core structure at high velocity. We will add finite temperature later as well, and eventually carry over this work into three dimensions and real potentials.

\vskip 15 pt
\noindent{\bf ACKNOWLEDGMENTS}
\vskip 10 pt

This project grew out of a collaboration with Jeff Tomasi, who originated our boundary condition method. It was supported by NSF grant number DMR 9873214, and was done using the Intel/NT Velocity Cluster at the Cornell Theory Center. We had helpful discussions with Tejs Vegge, Enrique Batista and Markus Rauscher.


\vskip 10 pt

\begin{thebibliography}{bb}
\bibitem{schiotz} cond-mat/9808211 J. Schiotz, T. Vegge, F. D. Di Tolla, K. W. Jacobsen \lq \lq Simulations of mechanics and structure of nanomaterials---from nanoscale to coarser scales"
\bibitem{Argon} M. P. Allen, D. J. Tildesley, \lq\lq Computer Simulation of Liquids", Oxford University Press (1987) p.21.
\bibitem{NEB} T. Rasmussen, K. W. Jacobsen, T. Leffers, O. B. Pedersen, S. G. Srinivasan and H. Jonsson, Phys. Rev. Lett. {\bf 79}, 3676 (1997).
\bibitem{xichen} X. Chen (private communication).
\bibitem{tomasi} J. Tomasi (private communication).
\bibitem{parrinello} M. Parrinello, A. Rahman, Phys. Rev. Lett. {\bf 45}, 1196 (1980); J. Appl. Phys.  {\bf 52}, 7182 (1981); J. Chem. Phys. {\bf 76}, 2662 (1982).
\bibitem{rates} P. H\"{a}nggi, P. Talkner, M. Borkovec, Rev. Mod. Phys. {\bf 62}, 251 (1990).
\end{thebibliography}
\end{document}